# Reversible thermoelectric nanomaterials

[1,2]T. E. Humphrey[*] and [3]H. Linke

[1]Engineering Physics, University of Wollongong, Wollongong 2522, Australia.
[2]Centre of Excellence for Advanced Silicon Photovoltaics and Photonics, University of New South Wales, N.S.W 2052, Australia.
[3]Materials Science Institute and Physics Department, University of Oregon, Eugene OR 97403-1274, U.S.A.

Irreversible effects in thermoelectric materials limit their efficiency and economy for applications in power generation and refrigeration. While electron transport is unavoidably irreversible in bulk materials, here we derive conditions under which reversible diffusive electron transport can be achieved in nanostructured thermoelectric materials via the same physical mechanism utilized in quantum optical heat engines. Our results may provide a physical explanation for the very high efficiencies recently reported for nanostructured thermoelectric materials such as quantum-dot superlattices.

The textbook example for heat engines are *cyclic* engines, such as the well-known Carnot, Otto and Brayton cycles, which may be used to model steam turbines or gasoline engines [1]. Cyclic heat engines transfer heat between at least two reservoirs via a working gas that moves through a number of quasi-equilibrium states. Reversibility may be achieved when the cycle progresses infinitely slowly [1].

Less well-known is a second distinct type, which we here denote as *particle exchange* (PE) heat engines. These transfer heat between at least two reservoirs via the continuous exchange of particles in a finite energy range in the presence of a field against which work is done. Reversibility is achieved when particles are transmitted only at the energy where the occupation of states in the reservoirs is equal [2-4]. Most PE heat engines studied previously have been discrete, with particles moving elastically between two reservoirs only, and include the three-level amplifier [2,5-6] and ballistic electron heat engines [3,7-8], which may be used to model lasers and thermionic devices, as well as solar cells and LEDs [4].

Nanostructured thermoelectric materials with delta-like electronic density of states (DOS) have recently been shown experimentally to have dramatically higher efficiencies than their bulk counterparts [9-11]. Here we model thermoelectric nanomaterials as PE heat engines in which electrons in a narrow energy band move diffusively through a material with a continuous spatial variation in temperature and electrochemical potential. We derive conditions under which thermoelectric nanomaterials can operate reversibly, so challenging a long held view that thermoelectric devices are inherently irreversible heat engines [12]. Our results have significant practical application in the design and optimization of new thermoelectric nanomaterials with ultra-high limiting efficiencies.

The efficiency of any heat engine is bounded above by the Carnot limit, which can only be achieved in systems infinitesimally close to an equilibrium state. An electronic system in equilibrium is characterized by a spatially invariant occupation of states given by the Fermi-Dirac (FD) distribution function,

$$f_{FD} = \left[\exp(s^{+1}(\mathbf{r})/k)+1\right]^{-1}, \quad (1)$$

where $s^{+1}(E,\mathbf{r})=[E-\mu(\mathbf{r})]/T(\mathbf{r})$ corresponds to the entropy change in the system if one electron with energy $E$ is added to the system at the spatial coordinate $\mathbf{r}$. A spatially invariant probability of occupation of available electronic states can be achieved in three ways, two of which are well known, while the third, a continuous form of the result for discrete PE heat engines, is pointed out here:

Firstly, a state of global equilibrium is attained when the electrochemical potential $\mu(\mathbf{r})$ and the temperature $T(\mathbf{r})$ are both constant as a function of $\mathbf{r}$. This corresponds to the textbook definition of an equilibrated electronic system.

Secondly, equilibrium can be approached when the only available electronic states in the system are at very high energies (for instance in an intrinsic semiconductor with bandgap, $E_g \to \infty$) where occupation tends to a spatially constant value of zero, irrespective of finite gradients in $\mu(\mathbf{r})$ and $T(\mathbf{r})$.

Here, we identify a third way in which equilibrium can be approached in continuous electronic systems, which we denote 'energy-specific equilibrium' [13]. We consider a material in which (i) the DOS for electrons is a δ-function at $E_0$, and in which (ii) $\mu(\mathbf{r})$ and $T(\mathbf{r})$ vary across the system in such a way that $s^{+1}(E_0,\mathbf{r})$ is spatially invariant. This means that the population of electron states at the specific energy $E_0$ is the same throughout the material (see Fig. 1 and 2c). Under these conditions the entire electronic system is in equilibrium, in spite of the thermal and potential gradients.

In the following we show how energy-specific equilibrium can be implemented in order to achieve reversible electron transport. We note that the rate of entropy production per unit volume, $\dot{s}$, in the material due to the movement of electrons in response to temperature and electrochemical potential gradients is [1]

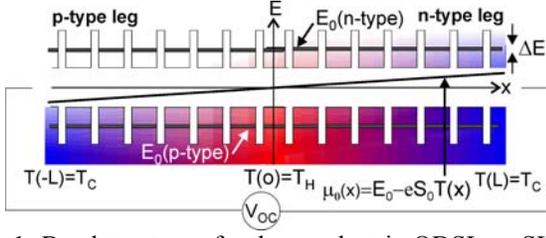

FIG. 1. Bandstructure of a thermoelectric QDSL or SLNW as used in our numerical model. Narrow minibands with width $\Delta E$ for electrons (holes) are located at energy $E_0$ (-$E_0$). The doping level varies across the material to ensure that $\mu(x)$ satisfies Eq. 3. Note that for simplicity, band-bending near the metallic contacts has not been considered.

$$\dot{s} = \nabla\left(\frac{1}{T}\right)\cdot\left(J_q + \mu J_n\right) + \nabla\left(\frac{\mu}{T}\right)\cdot J_n \qquad (2)$$

where $J_n$ is the number flux of electrons in the direction of decreasing temperature and $J_q$ the heat flux due to electrons, and where $\mu$, $T$ and $J_q$ depend on $\mathbf{r}$. In the limit that only electrons with energy $E_0$ are transmitted, the heat current in the material is given by $J_q(\mathbf{r}) = J_n[E_0 - \mu(\mathbf{r})]$. To find the conditions under which electrons move through the material without increasing the entropy of the system (reversible transport) we set $\dot{s}(E_0) = 0$ to obtain a differential equation which may be solved by integration with respect to $\mathbf{r}$ to find $E_0 = \mu(\mathbf{r}) + \Omega/T(\mathbf{r})$. Here $\Omega$ is an integration constant, evaluated using the boundary condition $\mu_C - \Omega T_C = \mu_H - \Omega T_H$, where the subscripts $_C$ and $_H$ refer to the cold and hot extremes of the system respectively. At open circuit, this yields $\Omega = -eV_{OC}/\Delta T$, where $eV_{OC} = \mu_C - \mu_H$, $\Delta T = (T_H - T_C)$ and where $\Delta T$ and $\mu_C$ (or alternatively $\mu_H$) are freely chosen parameters. We then arrive at our main result, namely an expression for a spatially varying chemical potential that ensures that electron transport at $E_0$ is reversible at open circuit:

$$\mu_0(\mathbf{r}) = E_0 - eV_{OC}[T(\mathbf{r})/\Delta T] \qquad (3)$$

(The subscript 0 refers to the state of energy-specific equilibrium). To clarify the physics involved in Eq. 4 we note that

$$S_0 \equiv V_{OC}/\Delta T = [E_0 - \mu_0(\mathbf{r})]/eT(\mathbf{r}) \qquad (4)$$

is the thermopower (Seebeck coefficient) corresponding to energy-specific equilibrium, which can be physically interpreted as the entropy carried by one Ampere of current [1] (that is, $eS_0 = s^{+1}(E_0)$). As $S_0$ is spatially invariant, there is no entropy increase when an electron with energy $E_0$ moves through the material, and therefore no thermodynamically spontaneous direction for current to flow, despite the finite thermal and electrical potential gradients, confirming that the electronic system is in equilibrium.

In practice, reversible electron transport can be approached by (i) creating a nanostructured material, such as a quantum dot superlattice (QDSL) [10] or a superlattice nanowire (SLNW) [11,14] with a DOS for mobile electrons that is sharply peaked at one energy $E_0$, and that is (ii) inhomogeneously doped such that Eq. 4 is fulfilled (Fig. 1). The extent to which a nanomaterial conforms to Eq. 4 could be tested by checking for spatial invariance of the Seebeck coefficient, for instance using the scanning probe technique recently developed by Lyeo et al. [15].

A material with a delta-function DOS doped according to Eq. 3 will have an 'electronic' efficiency (efficiency in the absence of phonon heat leaks) approaching the Carnot limit. To show this, we note that the heat flux withdrawn from the hot end of the system by electrons is $|J_q(T_H)| = (E_0 - \mu_H)|J_n|$ (Eq. 3). At open circuit the power is $P \equiv eV_{OC}|J_n| = eS_0(T_H - T_C)|J_n|$, giving an electronic efficiency $\eta_{PG} \equiv P/J_q(T_H) = (1 - T_C/T_H)$, which is the Carnot limit for power generation. Similarly, if the system is operated in reverse ($J_n \to -J_n$) as a refrigerator, then the coefficient of performance can be shown to be equal to the Carnot limit, $\eta_R \equiv J_q(T_C)/P = [T_C/(T_H - T_C)]$.

Our result also provides an explanation for the larger than expected thermopower of nanomaterials with delta-like DOS such as quantum dots [16] and QDSL [10], the physical origin of which has not to date been clear [17]. $S_0$ is in fact the theoretical upper bound upon the Seebeck coefficient $S$ for a particular choice of $\mu_C$ and $\Delta T$ (dropped across an infinitesimal length of material). To show this, we evaluate $\mu(\mathbf{r})$ and $T(\mathbf{r})$ in Eq. 4 at the hot end of the material to obtain $S_0\Delta T = (E_0 - \mu_H)(1 - T_C/T_H)$. This means that $S = S_0$ when the Carnot fraction, $(1-T_C/T_H)$, of heat removed by each electron with energy $E_0$ from the hot reservoir, $(E_0-\mu_H)$, is converted to useful work $S_0\Delta T = eV_{OC}$. $S_0$ therefore represents a theoretical maximum, as $S > S_0$ would imply an efficiency greater than the Carnot limit.

It is important to note that the presence of inelastic scattering in thermoelectric materials does not affect these results. Inelastic scattering processes produce heating via the relaxation of carriers from a non-equilibrium occupation of states to an occupation given by a FD distribution with the appropriate local value of $\mu(\mathbf{r})$ and $T(\mathbf{r})$. Since in our system states at $E_0$ are occupied with the same probability throughout the material (see Fig 2c), inelastic processes that scatter individual electrons into these states are just as likely as processes that scatter electrons out of these states, so there is no net exchange of energy between free electrons and the lattice due to the movement of electrons with energy $E_0$.

If the DOS is finite in some range around $E_0$, electrons occupying states at energies above (below) $E_0$ are, on average, scattered to lower (higher) energy states, depending upon the difference between the local probability of occupation and that of the arriving electrons. This process results in local heating (cooling) of the lattice.

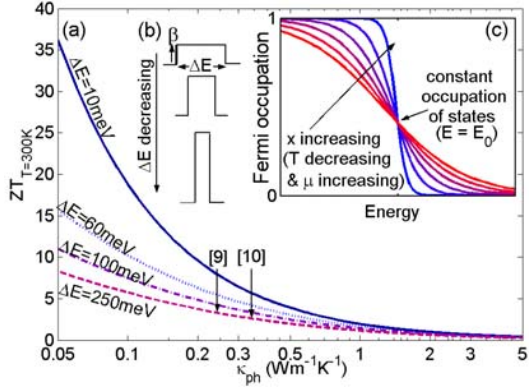

FIG 2. (a) $ZT$ at $T_C = 300$K as a function of $\kappa_{ph}$ for different $\Delta E$ as indicated. Marked [9] and [10] are experimental values for $\kappa_{ph}$ obtained at 300K in the nanomaterials described in the corresponding references. (b) Schematic of the variation of the magnitude of $\beta$ with decreasing $\Delta E$ resulting from the scaling operation described in the caption to Fig. 3. (c) Fermi occupation function at several places along the x-axis in the material shown in Fig. 1.

Crucially, however, the nearer these electrons are in energy to $E_0$, the smaller the heating (cooling) effect, due to the small variation in the occupation of states between adjacent regions of the material.

To quantify the advantage of using a delta-like DOS and inhomogeneous doping according to Eq. 4, we now numerically characterize a nanomaterial in which we vary the DOS from delta-like to bulk-like. We assume a finite lattice thermal conductivity ($\kappa_{ph} \neq 0$) and a single miniband of width $\Delta E$ as shown in Fig. 1. We use the Boltzmann transport equation under the relaxation-time approximation [18]. Electrical conductivity, $\sigma$, thermal conductivity due to electrons, $\kappa_{el}$, and $S$, and can all be expressed as a function of the integral [18]

$$K_\alpha = \int \beta(E)(E-\mu)^\alpha \left(-\frac{df}{dE}\right) dE \quad (5)$$

where $\beta(E) = D(E)\tau(E)v(E)$, $\tau(E)$ is the electron relaxation time, $v(E)$ is the electron group velocity, $D(E)$ is the DOS and where $\alpha = 0$, 1 or 2. Then $\sigma = e^2 K_0$, $S = -K_1/(eTK_0)$, $\kappa_{el} = (K_2 - K_1^2/K_0)/T$ and the dimensionless figure of merit $ZT = T\sigma S^2/(\kappa_{el}+\kappa_{ph})$. We use $\kappa_{ph} = 0.33$ Wm$^{-1}$K$^{-1}$, the value measured by Harman et al. for PbSeTe/PbTe QDSL [10]. For simplicity and transparency in the numerical results, we assume that $\beta(E) = \beta$ is constant over the energy range $\Delta E$, a reasonable assumption for the small values of $\Delta E$ in which we are primarily interested. To isolate the effect upon $ZT$ of changing $\Delta E$ from effects due to changing the overall *number* of states, we scale the magnitude of $\beta$ with $\Delta E$ such that for all values of $\Delta E$, $\sigma = 5\times 10^5$ $\Omega^{-1}$m$^{-1}$ at $\mu_c = E_0$ (see Fig. 2b and the figure caption for Fig. 3 for details). This choice means that at the values of $\mu_C$ for which $ZT$ is optimized at $T_C = 300$K, we obtain $\sigma \approx 6\times 10^4$ $\Omega^{-1}$m$^{-1}$ for $\Delta E \approx 200$meV, the same conductivity as measured in [10] (the energy spectrum of the DOS in [10] is not known). In addition, we obtain $\sigma \approx 1\times 10^5$ $\Omega^{-1}$m$^{-1}$ for an optimized $ZT$ at $\Delta E \approx 60$ meV, which is similar to the value obtained by numerical modeling in [11] for a PbSe/PbS SLNW with a 60 meV wide miniband.

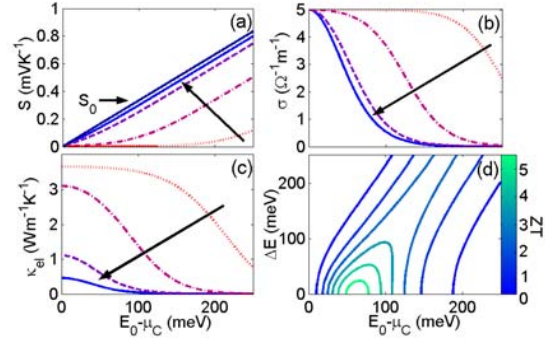

FIG. 3. Numerically calculated thermoelectric parameters as a function of $(E_0 - \mu_C)$ for $\Delta E = 500$meV, 250meV, 100meV, and 60meV (arrows indicate decreasing $\Delta E$), for $T(L) = 300$K and $\kappa_{ph} = 0.33$ Wm$^{-1}$K$^{-1}$. For each $\Delta E$ we used the value $\beta = 5\times 10^5/ e^2 K_0(\Delta E, \mu_c = E_0)$ to calculate $K_0$, $K_1$ and $K_2$ for $\mu_c \neq E_0$. See main text for further model details. (a) Seebeck coefficient. (b) Electrical conductivity. (c) Electronic thermal conductivity. (d) Contour map of ZT. Figure (a) is antisymmetric and figures (b – d) mirror symmetric with respect to $(E_0 - \mu_C) = 0$ (not shown).

Fig. 3a shows $S$ as a function of $(E_0-\mu_C)$ at the position $x = L$ for different $\Delta E$. As $\Delta E$ decreases the system approaches energy-specific equilibrium and $S$ approaches $S_0$, the theoretical maximum Seebeck coefficient given by Eq. 4.

Fig. 3b and 3c show $\sigma$ and $\kappa_{el}$, respectively, as a function of $(E_0-\mu_C)$. Note that although $\sigma$ at $\mu_C = E_0$ is kept constant with decreasing $\Delta E$, $\kappa_{el}$ at $\mu_C = E_0$ decreases, that is, $K_1^2/K_0 \rightarrow K_2$ [19]. An important implication of this result is that the Wiedemann-Franz law [18], often used by experimentalists to calculate $\kappa_{el}$ from the electrical conductivity [9-10], is not applicable to nanomaterials with delta-like DOS. Crucially, this result also means that thermoelectric nanomaterials are not subject to the conundrum that limits $ZT$ of bulk thermoelectrics: the interrelationship of $S$, $\sigma$ and $\kappa_{el}$.

To emphasize this point, we show in Fig. 3d that for small $\Delta E$ the figure of merit is optimal at values of $(E_0-\mu_C)$ [20] where $S$ (Fig. 3a) and $\kappa_{el}$ (Fig. 3c) are both relatively small. In other words, the best strategy to maximize $ZT$ in nanomaterials is to minimize $\kappa_{el}$ leaving $ZT$ limited only by finite $\kappa_{ph}$. The fact that this is possible without at the same time decreasing $\sigma$ illustrates a fundamental difference between the thermodynamics of nanostructured and bulk thermoelectric materials.

Using the technique outlined in [21], the difference between inhomogeneous doping according to Eq. 3 ($\mu(x) = \mu_0(x)$) and homogeneous doping ($\mu(x) = \mu_0(L/2)$) can be quantified. We find [22] that inhomogeneous doping can increase the maximum efficiency by 10% (corresponding to a doubling of $ZT$ at 300K for $\Delta E =10$meV, $T_H = 800$K, $T_C = 300$K and $\kappa^{ph} = 0.5$ Wm$^{-1}$K$^{-1}$) and increase the maximum power by up to 60%. The physics behind this improvement is now clear; inhomogeneous doping increases $ZT$ as it brings thermoelectric materials closer to a state of energy-specific equilibrium.

Figure 2a shows $ZT$ at 300K as a function of $\kappa_{ph}$, demonstrating that while a decrease in $\kappa_{ph}$ is always beneficial to the figure of merit, a decrease in $\kappa_{ph}$ *combined* with a decrease in $\Delta E$ results in spectacular increases in $ZT$. The development of nanomaterials with a $\kappa_{ph}$ 20% lower than current state-of-the-art materials and optimized, delta-like DOS could result in $ZT = 10$ at $T = 300$K, far exceeding the range of $ZT > 5$ required for economical adoption of thermoelectric technology for main stream refrigeration and power generation.

Finally, we note the finite coherence length of electrons, which places a lower limit on the width of DOS peaks resulting from quantum confinement, will also in principle limit the efficiency of thermoelectric nanomaterials. However, as all heat engines are in reality operated at finite power, away from maximum efficiency, quantum efficiency limits are not expected to be a practical design issue.

*Email address: tammy.humphrey@unsw.edu.au; Web page: www.humphrey.id.au

1. Herbert B. Callen, *Thermodynamics and an introduction to Thermostatics* (Wiley, 1985).
2. H.E.D. Scovil and E.O. Schulz-DuBois, Phys. Rev. Lett. **2**, 262 (1959).
3. T.E. Humphrey, R. Newbury, R.P. Taylor, and H. Linke, Phys. Rev. Lett. **89**, 116801 (2002).
4. T. E. Humphrey, H. Linke, cond-mat/0407508 (2004).
5. T. Feldmann and R. Kosloff, Phys. Rev. E **61,** 4774 (2000).
6. M. O. Scully, Phys. Rev. Lett. **88,** 050602 (2002).
7. T. E. Humphrey, Ph.D. thesis, University of New South Wales, 2003. http://adt.caul.edu.au/
8. T. E. Humphrey and H. Linke, submitted to Appl. Phys. Lett. (2004). cond-mat/0401377
9. R. Venkatasubramanian, E. Siivola, T. Colpitts, B. O'Quinn, Nature **413**, 597 (2001).
10. T. C. Harman, P. J. Taylor, M. P. Walsh and B. E. LaForge, Science **297**, 2229 (2002).
11. Yu-Ming Lin, M. S. Dresselhaus, Phys. Rev. B **68**, 075304 (2003).
12. H. Littman and B. Davidson, J. Appl. Phys. **32**, 217 (1961).
13. Our conclusions are equally valid for continuous or discrete systems in which particles obey Bose-Einstein or Maxwell-Boltzmann statistics.
14. High-yield bulk fabrication of SNLW should be possible using heterostructured whiskers as described in: C. Thelander et al., Appl. Phys. Lett. **83** 2052 (2003).
15. Lyeo et al., Science **303**, 816 (2004).
16. C. W. J. Beenakker and A. A. M. Staring, Phys. Rev. B **46**, 9667 (1992).
17. G. Chen et al., Int. Mat. Rev. **48**, 1 (2003).
18. N. W. Ashcroft and N. D. Mermin, *Solid State Physics* (Saunders College Publishing, 1976).
19. G. D. Mahan and J. O. Sofo, Proc. Nat. Acad. Sci. **93**, 7436 (1996).
20. In (*19*) it is shown that the figure of merit of a thermoelectric material with a δ-function DOS is optimised at $(E_0-\mu_C) = 2.4kT$ ($\approx 62$meV at 300K).
21. G. D. Mahan, J. Appl. Phys. **70**, 4551 (1991).
22. T. E. Humphrey and H. Linke, Presented at the International Thermoelectrics conference 2004. (cond-mat/0407506).
23. TH is supported by the ARC and SERDF, and HL by a CAREER grant from the NSF.